\documentclass[twocolumn,prl,aps,psfig,showpacs,preprintnumbers]{revtex4}
\usepackage{graphicx}
\usepackage{epstopdf}
\usepackage{float}
\usepackage{color}
\usepackage{amsmath}

\begin{document}
\title{NMR evidence for static local nematicity and its cooperative interplay with low-energy magnetic fluctuations in FeSe under pressure}

\author{P.~Wiecki, M.~Nandi, A.~E.~B\"{o}hmer, S.~L.~Bud'ko, P.~C.~Canfield, Y.~Furukawa}
\affiliation{Ames Laboratory, U.S. DOE and Department of Physics and Astronomy, Iowa State University, Ames, Iowa  50011  USA}
\date{\today}

\begin{abstract}
    We present $^{77}$Se-NMR measurements on single-crystalline FeSe under pressures up to 2 GPa. 
    Based on the observation of the splitting and broadening of the NMR spectrum due to structural twin domains, we discovered that static, local nematic ordering exists well above the bulk nematic ordering temperature, $T_{\rm s}$. 
    The static, local nematic order and the low-energy stripe-type antiferromagnetic spin fluctuations, as revealed by 
    NMR spin-lattice relaxation rate measurements, are both insensitive to pressure application. 
    These NMR results provide clear evidence for the microscopic cooperation between magnetism and local nematicity in FeSe. 

\end{abstract}

\maketitle
 
      Much attention in recent research on iron-based superconductivity (SC) has been paid to understanding the nature of the electronic nematic phase, which breaks rotational symmetry while preserving time-reversal symmetry \cite{Fernandes2012,Fernandes2014}.
   In the archetypical ``122" compounds AFe$_2$As$_2$ (A=Ca, Sr, Ba) \cite{Canfield2010,Johnston2010}, 
   the nematic phase is closely tied to the stripe-type antiferromagnetic (AFM) phase in the phase diagram, suggesting a magnetic origin for the nematic state. 
        Among the Fe-based SCs, FeSe is known to be an exception. 
   At ambient pressure, FeSe undergoes a transition to the nematic phase at a bulk structural phase transition temperature $T_{\rm s}\sim90$~K, as well as to SC below $T_{\rm c}\sim8$~K, but has no stripe-type AFM ordered phase.
   Under pressure ($p$), $T_{\rm s}$ is suppressed to $\sim$20~K at $p$ $\sim$1.7 GPa \cite{Miyoshi2014,Terashima2015,Kothapalli2016}
   and an AFM ordered state emerges above $\sim$0.8 GPa \cite{Bendele2010,Bendele2012,Kaluarachchi2016,Terashima2016}. 
    In addition, $T_{\rm c}$ is enhanced from 8 K at ambient pressure to $\sim$37 K at $p\sim6$ GPa \cite{Sun2016}. 
     The decrease of $T_{\rm s}(p)$ and increase of $T_{\rm N}(p)$ under pressure 
     suggests competition between nematic and magnetic orders. 
    Furthermore, NMR measurements \cite{Bohmer2015,Baek2015} showed Korringa behavior above $T_{\rm s}$, consistent with an uncorrelated Fermi liquid, while AFM spin fluctuations (SFs) were found to be strongly enhanced only below $T_{\rm s}$.
    These observations suggested that SFs are not the driver for nematic order and therefore pointed to an orbital mechanism for the nematicity \cite{Baek2015}. 
An orbital mechanism was also suggested by Raman spectroscopy \cite{Massat2016}.

     In contrast, several recent studies
    have suggested cooperation between nematicity and magnetism in FeSe. 
   High-energy x-ray diffraction measurements \cite{Kothapalli2016} found that the orthorhombic distortion is enhanced in the magnetic state at $p=1.5$ GPa.  
    Furthermore, above 1.7 GPa $T_{\rm s}(p)$ and $T_{\rm N}(p)$ were found to coincide as a simultaneous first-order magneto-structural transition.
    These observations are consistent with a spin-driven mechanism for nematic order in FeSe. 
     Similarly, inelastic neutron scattering (INS) measurements at ambient pressure \cite{Wang2015,QWang2016} showed that commensurate stripe-type AFM SFs are in fact present well above $T_{\rm s}$, which could possibly drive the nematic transition. 
     These SFs were not  seen by NMR \cite{Bohmer2015,Baek2015} due to a spin gap above $\sim90$ K. 
In addition, $^{77}$Se-NMR data under pressure \cite{Wang2016} revealed a first-order transition to a stripe-type magnetic ordered state, and 
suggested a magnetic driven nematicity.
     Therefore, the origin of nematicity in FeSe is still under intense debate, motivating further study of the microscopic properties of the nematic state
     in FeSe.

    Here, we present $^{77}$Se-NMR measurements on FeSe under pressures up to 2 GPa, focusing our attention on the local, microscopic properties of the paramagnetic and nematic phases. 
   We found clear evidence that a static, local nematic ordering exists well above $T_{\rm s}$.
   Both the local nematic order and the low-energy stripe-type antiferromagnetic spin fluctuations, are found to be robust against pressure application,
    providing clear evidence for the microscopic cooperation between magnetism and local nematicity in FeSe.


\begin{figure*}[t]
\centering
\includegraphics[width=\textwidth]{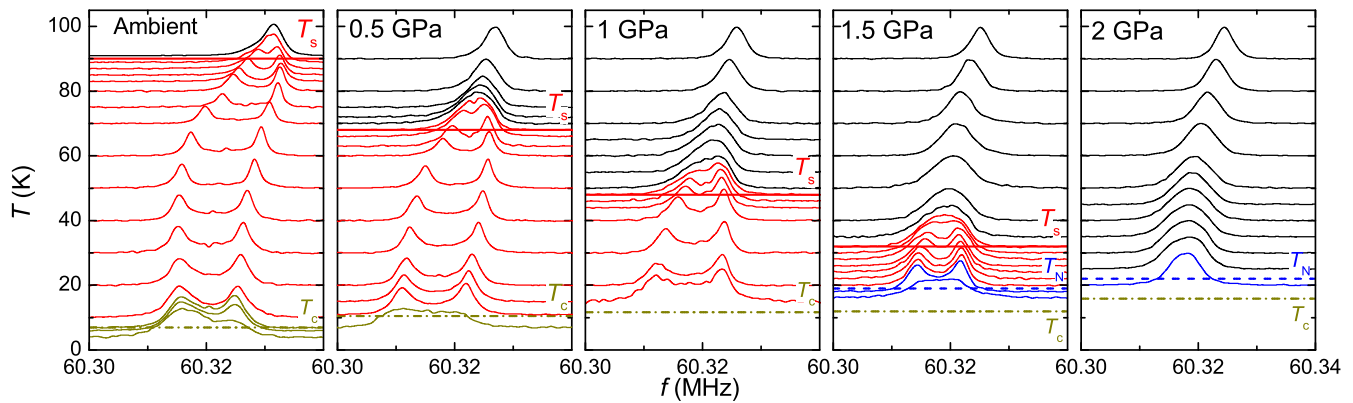}
\caption{Representative NMR spectra at indicated $T$ for all measured pressures. 
The two peaks arise from inequivalent nematic domains with $H\|a$ and $H\|b$.
The solid red line is the bulk $T_{\rm s}$, as determined by a kink in the NMR shift (see Fig. \ref{fig:FWHM}). 
The dashed blue line is $T_{\rm N}$ and the dot-dash dark yellow line is $T_{\rm c}$. 
$T_{\rm N}$ and $T_{\rm c}$ were determined from data shown in  shown in \cite{supp}. 
The colors of the spectra correspond to different phases: black for the paramagnetic state, dark yellow for below $T_{\rm c}$, blue for below $T_{\rm N}$, and red for  
the bulk nematic ordered state.  
  }
\label{fig:spectrum}
\end{figure*}

    $^{77}$Se-NMR ($I=1/2$; $\gamma/2\pi=8.118$ MHz/T) spectra have been measured on a single crystal (24 mg) of FeSe  in the temperature ($T$) range of 4--300 K with a fixed field of $H=7.4089$ T applied along the [110] direction in the high-$T$ tetragonal phase. 
    The crystal was grown using chemical vapor transport as outlined in Ref. \cite{Bohmer2016}.  
     At room temperature,  the spectra are very narrow with the full-width-at-half-maximum (FWHM)
      reaching as low as $\sim1.5$ kHz, which is half of 3 kHz reported previously \cite{Wang2016}, indicative of the high quality of our single crystal.  
    Typical NMR spectra below 100 K for all measured pressures are shown in Fig.~\ref{fig:spectrum}. 
    At ambient pressure, a clear splitting of the spectrum was observed  in the orthorhombic structural phase below $T_{\rm s}$, consistent with previous data \cite{Bohmer2015,Baek2015,Baek2016,Wang2016}. 
    The spectral splitting arises from the presence of two types of nematic domains in the twinned sample, one of which experiences $H\|a$ axis and the other $H\|b$ axis, combined with the anisotropy of the in-plane Knight shift ($K_a$ and $K_b$) in the nematic ordered phase \cite{Bohmer2015,Baek2015}.
    The difference of the Knight shift  $\Delta K = |K_a-K_b|$ is, therefore,  a measure of the local microscopic nematic order parameter \cite{Baek2015,Baek2016}.
    Under pressure \cite{supp}, we observed similar clear splittings of the spectra below the bulk $T_{\rm s}$ as shown in  Fig. \ref{fig:spectrum}. 
    However,  we found that the splitting of the spectrum exists even above $T_{\rm s}$ at all measured pressures.
   This was not reported in the previous NMR study \cite{Wang2016}.  
    A similar splitting of the spectrum above the bulk $T_{\rm s}$ was reported at ambient pressure due to random local strains produced by gluing of the crystal \cite{Baek2016}. 
     The asymmetric spectra observed for $T>T_{\rm s}$ originates from the difference in the FWHM of the lower- and higher-frequency peaks. 
     This provides evidence of the existence of the two peaks above $T_{\rm s}$, although the origin of the different FWHM of the two lines is not clear at present.
 The existence of two peaks above $T_{\rm s}$ under pressure is also shown by the $T$ and $p$ dependence of the coefficient of determination ($R^2$) of a single-peak fit shown in  Supplemental Materials (SM) \cite{supp}.

\begin{figure*}[t]
\centering
\includegraphics[width=\textwidth]{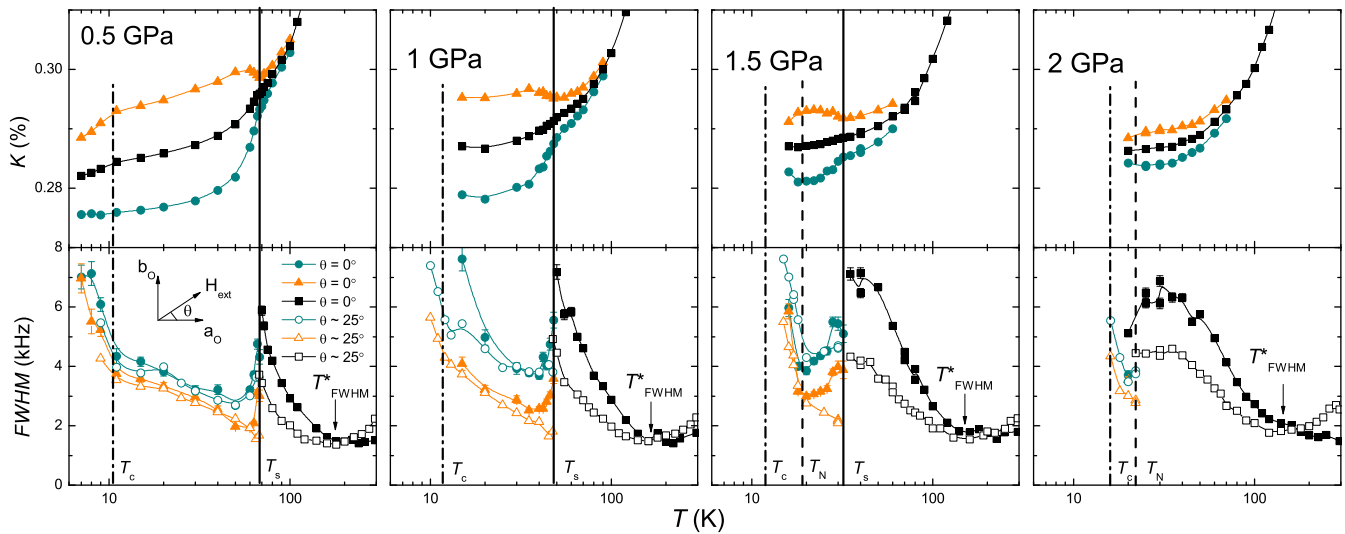}
\caption{
Upper panels: NMR shift $K$ as a function of $T$ as obtained from a two-Lorentzian fit for indicated pressures. 
Orange triangles and teal circles represent $K_a$ and $K_b$, while the black squares are the average of the two, $K_{\rm{avg}}$. 
The black vertical lines  indicate the corresponding bulk $T_{\rm s}$ for each pressure. 
 $T_{\rm c}$ and $T_{\rm N}$ for different pressures are also shown by the vertical broken lines.
 Lower panels: Full-width-at-half-maximum (FWHM) of NMR spectral peaks for two $ab$ plane orientations: $\theta=0^\circ$ (filled symbols)
 and $\theta\sim25^\circ$ (open symbols). 
 Below $T_s$, the FWHM of each of the two peaks is shown separately. The low-frequency peak (teal) has consistently greater FWHM than the high-frequency peak (orange). Above $T_s$ the FWHM of a single-peak model is shown (black). Arrows denote $T^*_{\rm{FWHM}}$.
}
\label{fig:FWHM}
\end{figure*}

        In order to extract $\Delta K$, we have fit the spectrum to a sum of two Lorentzian peaks. 
     From the fitting, we determined the position of each peak, providing the $T$ and $p$ dependence of $K_a$ and $K_b$ as shown by orange triangles and teal circles in the upper panels of Fig. \ref{fig:FWHM}. 
     Note that the NMR data alone do not determine which of the two peaks corresponds to $K_a$ \cite{Bohmer2015,Baek2015}.
     Also displayed is  the average value $K_{\rm{avg}}$ = $\frac{1}{2}$($K_a$ + $K_b$), shown by black squares. 
     $K_{\rm{avg}}$ decreases monotonically with decreasing $T$. 
    The bulk $T_{\rm s}$ is  identified by kinks in $K_a$, $K_b$ and $K_{\rm{avg}}$ as can be seen in the upper panel of  Figs. \ref{fig:FWHM}. 
    The observed  $T_{\rm s}$ agree well with values reported previously \cite{Terashima2016,Kaluarachchi2016,Kothapalli2016}.

\begin{figure}[t]
\centering
\includegraphics[width=\columnwidth]{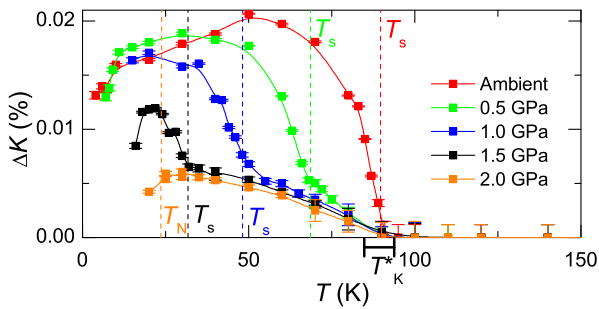}
\caption{  $T$ dependence of $\Delta K$ for the indicated pressures. 
$\Delta K$ is a measure of the local nematic order parameter. 
The dashed vertical lines indicate the bulk $T_{\rm s}$. 
}
\label{fig:deltaK}
\end{figure}

    Figure \ref{fig:deltaK} shows the $T$ dependence of $\Delta K$ under different pressures, where the vertical lines indicate the corresponding bulk $T_{\rm s}$ for each pressure. 
    At ambient pressure, $\Delta K$ increases sharply below $T_{\rm s}$ and shows a broad peak near $\sim50$ K before decreasing at low $T$,
    consistent with the previous NMR results \cite{Bohmer2015,Baek2015}. 
    A peak near $\sim60$ K is also seen in the $T$  dependence of the resistivity anisotropy \cite{Tanatar2016}.
    As seen from Fig. \ref{fig:deltaK}, $\Delta K$ remains non-zero within our experimental uncertainty above $T_{\rm s}$ up to a temperature we define as $T^*_K$.
    At ambient pressure, we find $T^*_K$ $\sim$ $T_{\rm s}$. 
    Under pressure, on the other hand, it is clearly seen that $T^*_K$ exceeds $T_{\rm s}$. 
        It is also found that $T^*_K$ is nearly constant as a function of $p$, despite the decrease of $T_{\rm s}$.
          Given the fact that  recent x-ray diffraction measurements \cite{Kothapalli2016} indicated  that the bulk tetragonal symmetry of the crystal is broken only below $T_{\rm s}$, our NMR results imply that a short-range nematic order exists above $T_{\rm s}$ in the bulk tetragonal phase, 
          and is surprisingly resistant to pressure application.
            Since the NMR spectrum probes static electronic properties,  these results indicate that the local nematic short-range order is static at the NMR time scale ($\sim$MHz).
A similar local static nematic state has been observed in the BaFe$_2$(As$_{1-x}$P$_x$)$_2$ system \cite{Iye2015,Kasahara2012} in which  NMR spectrum measurements on the $x=0.04$ ($T_{\rm s}=120$ K) compound revealed the existence of nearly static nematic fluctuations up to 250 K \cite{Iye2015}.

Evidence for nematicity above $T_{\rm s}$ is also seen in the FWHM of the spectra (Fig. \ref{fig:FWHM} lower panels). 
   In the PM state, the FWHM displays a strong upturn at a pressure-dependent temperature $T^*_{\rm{FWHM}}$, indicated by black arrows in Fig. \ref{fig:FWHM}.
   Since $^{77}$Se has $I=1/2$, the broadening cannot be attributed to quadrupole effects. 
    In normal circumstances of magnetic broadening of NMR lines in a paramagnetic (PM) phase, the FWHM is expected to have the same $T$ dependence as the NMR shift $K$, which measures the uniform spin-susceptibility of the electrons. 
    In FeSe, we find that $K$ decreases monotonically with decreasing temperature \cite{supp}, consistent with \cite{Imai2009,Bohmer2015,Baek2015}. 
    The observed increase in the FWHM is therefore quite unexpected for a PM state, and cannot be due to normal magnetic broadening effects.

     To get further insight into the origins of the increase in FWHM we also measured the spectrum with the crystal rotated by $\theta\sim25^\circ$ 
away from tetragonal [110] within the $ab$ plane. 
     At ambient pressure, Baek \cite{Baek2015} has shown explicitly that $\Delta K$ below $T_{\rm s}$ vanishes at $\theta=45^\circ$, since then both types of domains experience symmetry-equivalent magnetic field directions. Indeed, we find that $\Delta K$ below $T_{\rm s}$ is much reduced at $\theta\sim25^\circ$ \cite{supp}.
     Remarkably, we find that the FWHM above $T_s$ is also drastically reduced at $\theta\sim25^\circ$.
    However, below $T_{\rm s}$ the FWHM of the two individual peaks shows no $ab$ plane orientation dependence.
    $T^*_{\rm{FWHM}}$ also has no $ab$ plane orientation dependence.
    These results, together with the asymmetric shape of the spectra described above, clearly
    indicate that the broadening above $T_{\rm s}$ is due to local nematicity and not local magnetism.
  We conclude that above $T_{\rm s}$ the NMR spectrum consists of two nematic peaks (of orientation independent FWHM) with a small, unresolved, orientation-dependent splitting. 
$T^*_{\rm{FWHM}}$ is understood as a crossover between magnetic- and nematic-dominated broadening. Local 
nematicity may therefore be present even above $T^*_{\rm{FWHM}}$ where the nematic splitting would be less than the magnetic broadening.

\begin{figure*}[t]
\centering
\includegraphics[width=\textwidth]{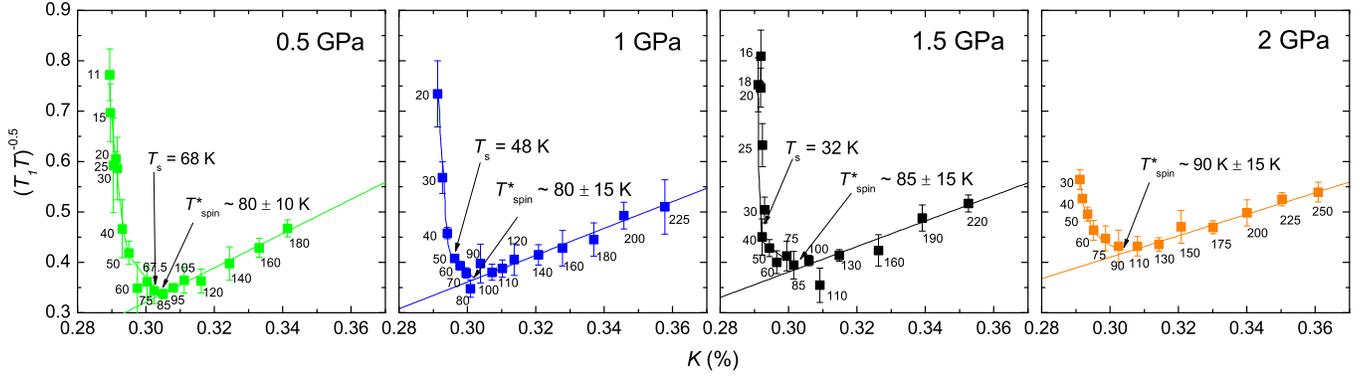}
\caption{$\sqrt{1/T_1T}$ versus $K(T)$ plot with $T$ as an implicit parameter for indicated pressures. 
$T^*_{\rm{spin}}$, the onset of low-energy spin fluctuations, is determined by the deviation of the data from high-$T$-linear behavior shown by solid lines (see text).
Bulk $T_{\rm s}$ is indicated for comparison. 
}
\label{fig:korringa}
\end{figure*}

\begin{figure}[tb]
\centering
\includegraphics[width=\columnwidth]{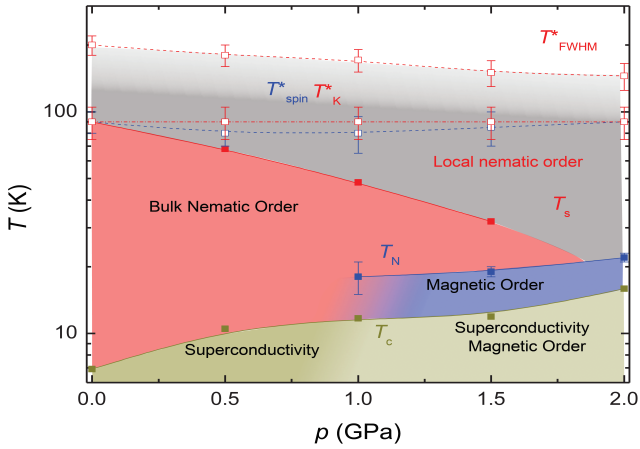}
\caption{Phase diagram of pressurized FeSe incorporating microscopic details of the paramagnetic phase as revealed by NMR. 
}
\label{fig:phase}
\end{figure}

     We now discuss the AFM SFs based on the $^{77}$Se spin-lattice relaxation rate  1/$T_1$ and $K$ data.
   For all pressures measured, as shown in SM \cite{supp}, 1/$T_1T$ shows a similar $T$ dependence in which 1/$T_1T$ decreases with decreasing $T$ from room temperature to around $T$ $\sim$ 80~K then increases, which indicates enhancements of  low-energy AFM  SFs at low $T$ \cite{Imai2009}. 
    Within a Fermi liquid picture, the spin part of the NMR shift $K_{\text{s}} (\propto \chi_{\text{spin}}$) is proportional to the density of states at the Fermi energy ${\cal D}(E_{\rm F})$, whereas    
    $1/T_1T$ is proportional to the square of ${\cal D}(E_{\rm F})$.
   Therefore, in order to examine electron correlation effects, it is useful to estimate the quantity $T_1TK_{\rm s}^2$ \cite{Moriya1963,Narath1968}. 
 The so-called Korringa ratio 
 $\alpha = \hbar\gamma_{\rm e}^2/(T_1TK_{\rm s}^24\pi k_{\rm B} \gamma_{\rm n}^2)$
 is unity for uncorrelated metals. 
    Here we plot $\sqrt{1/T_1T}$ vs. $K(T)$ with $T$ as an implicit parameter, for which a straight line is expected for the Korringa behavior. 
    Under ambient pressure, the Korringa behavior is observed above $T_{\rm s}$ and $\alpha$ is estimated to be $\sim$ 1, suggesting  no significant AFM correlations above $T_{\rm s}$. 
    On the other hand, below $T_{\rm s}$,  enhancements of AFM SFs are observed via the deviation of $\sqrt{1/T_1T}$ from the high-$T$ linearity \cite{Bohmer2015}. 

    The $\sqrt{1/T_1T}$ vs. $K(T)$ plots for all measured pressures are shown in Fig. \ref{fig:korringa} where, for reference, the $T$ for each point is indicated. 
     At 0.5 GPa, the $\sqrt{1/T_1T}$ vs $K(T)$ behavior is similar to the case for  ambient pressure, but one can see a deviation of $\sqrt{1/T_1T}$ from the high-$T$ linearity slightly above $T_{\rm s}$,  indicating that AFM SFs are enhanced slightly above $T_{\rm s}$.
         This effect is much more apparent at higher pressures. 
    We define $T_{\rm spin}^*$ as the temperature below which low-energy SFs are enhanced. At ambient pressure, $T_{\rm spin}^*\sim T_s$ \cite{Bohmer2015}.
     At 1 GPa, we find $T^*_{\rm{spin}}\sim80$ K which differs significantly from $T_{\rm s}=48$ K. 
    A similar behavior is also observed at 1.5 GPa with $T_{\rm s}$ =  32 K and $T_{\rm spin}^*\sim$ 85 K. 
     At 2 GPa, we find $T^*_{\rm{spin}}\sim90$ K.
   
As seen in the phase diagram of Fig. \ref{fig:phase}, $T^*_{\rm spin}$ is nearly pressure independent. 
This behavior is reminiscent of the 
robustness of $T^*_K$ (and $T^*_{\rm{FWHM}}$) to pressure application, suggesting a correlation between the local nematicity and low-energy magnetic fluctuations. 
While local nematicity is also present above $T^*_K$, its $\Delta K$ is too small to detect directly. 
It is possible that a corresponding small low-energy SF contribution
to $1/T_1T$ exists above $T^*_{\rm spin}$ which cannot be detected within experimental uncertainty.

    According to the INS measurements at ambient pressure \cite{Wang2015}, stripe-type AFM SFs exist above $T^*_{\rm spin}$, despite not being observed in our NMR measurements.
     Since NMR detects SFs in the very low-energy region (of order $\mu$eV) while INS probes mainly high-energy spin dynamics (of order meV), the AFM SFs must 
     have no spectral weight in the low-energy region which NMR can detect. 
     In fact, the INS measurements point out the existence of a spin gap of $\sim2.5$ meV at 110 K \cite{Wang2015}.
     The INS measurements also indicate that the spin gap is closed  below $T_{\rm s}$ at ambient pressure. 
     This picture is consistent with the NMR data at ambient pressure \cite{Bohmer2015}. 
     Since we continue to observe Korringa behavior above $T_{\rm spin}^*$ for all measured pressures, the high-$T$ spin gap which exists at ambient pressure remains present up to at least 2 GPa. 
        Therefore, $T^*_{\rm spin}(p)$ may be attributed to the closing of a spin gap.
  Since the argument for orbital-driven nematicity from the ambient pressure NMR data \cite{Baek2015,Bohmer2015} 
     is based on the lack of SFs above $T_s$, our observation of SFs above $T_s$ under pressure, combined with 
     the ambient pressure INS results, does not exclude the possibility of spin-driven nematic order. 
  Further studies are highly required to shed light on the nature of the spin gap in FeSe.

     In summary, from our measurements of the splitting and FWHM of $^{77}$Se-NMR spectra, 
     we find that a static, local nematic order exists above $T_{\rm s}$ in FeSe under pressure, which has not been detected in previous studies. 
          The local nematic order and the low-energy stripe-type antiferromagnetic spin fluctuations  
          are both nearly independent of pressure, suggesting a cooperation between the magnetic fluctuations and local nematicity in pressurized FeSe.
  
      The research was supported by the U.S. Department of Energy, Office of Basic Energy Sciences, Division of Materials Sciences and Engineering. Ames Laboratory is operated for the U.S. Department of Energy by Iowa State University under Contract No.~DE-AC02-07CH11358.
    
           {\it Note added.}\textemdash After submission of our manuscript, a similar NMR study, consistent with our results, was posted to the arXiv by Wang et al. \cite{Wang2017}.

\end{document}